\pdfoutput=1
\documentclass[11pt]{article}

\usepackage[margin=1in]{geometry}
\usepackage{times}
\usepackage{latexsym}

\usepackage[numbers]{natbib}

\usepackage{array}
\usepackage{multirow}
\usepackage{longtable}

\usepackage[T1]{fontenc}
\usepackage[utf8]{inputenc}
\usepackage{microtype}
\usepackage{inconsolata}
\usepackage{xspace}

\usepackage{graphicx}

\usepackage{xcolor}

\newcommand{\our}{ARBITER\xspace}
\usepackage{hyperref}
\hypersetup{
    colorlinks=true,
    linkcolor=blue,
    filecolor=magenta,      
    urlcolor=cyan,
    citecolor=blue,
}

\title{\our: AI-Driven Filtering for Role-Based Access Control}

\author{
\small
\begin{tabular}[t]{c}
Michele Lorenzo \\
Reply Spike \\
Turin, Italy \\
\texttt{mi.lorenzo@reply.it}
\end{tabular}
\hfill
\begin{tabular}[t]{c}
Idilio Drago \\
University of Turin \\
Turin, Italy \\
\texttt{idilio.drago@unito.it}
\end{tabular}
\hfill
\begin{tabular}[t]{c}
Dario Salvadori \\
Reply Spike \\
Milan, Italy \\
\texttt{d.salvadori@reply.it}
\end{tabular}
\hfill
\begin{tabular}[t]{c}
Fabio Romolo Vayr \\
Reply Spike \\
Turin, Italy \\
\texttt{f.vayr@reply.it}
\end{tabular}
}

\date{}

\begin{document}
\maketitle

\begin{abstract}
Role-Based Access Control (RBAC) struggles to adapt to dynamic enterprise environments with documents that contain information that cannot be disclosed to specific user groups. As these documents are used by LLM-driven systems (e.g., in RAG) the problem is exacerbated as LLMs can leak sensitive data due to prompt truncation, classification errors, or loss of system context. We introduce \our, a system designed to provide RBAC in RAG systems. \our implements layered input/output validation, role-aware retrieval, and post-generation fact-checking. Unlike traditional RBAC approaches that rely on fine-tuned classifiers, \our uses LLMs operating in few-shot settings with prompt-based steering for rapid deployment and role updates. We evaluate the approach on 389 queries using a synthetic dataset. Experimental results show 85\% accuracy and 89\% F1-score in query filtering, close to traditional RBAC solutions. Results suggest that practical RBAC deployment on RAG systems is approaching the maturity level needed for dynamic enterprise environments.
\end{abstract}

\section{Introduction}

Enterprises are rapidly embracing generative AI technologies, deploying them across local and cloud infrastructures to enhance productivity, streamline workflows, and support decision-making~\citep{RASHID2024100277}. Among these, Retrieval-Augmented Generation (RAG) has emerged as a powerful paradigm, allowing Large Language Models (LLMs) to access external knowledge sources and produce more accurate, contextually grounded outputs~\citep{lewis2020retrieval}. However, the integration of such capabilities in enterprise environments introduces security and access control challenges~\citep{huang2024genai}. 

A key problem is the risk of unauthorized access to sensitive data during the retrieval process. Generative systems may expose or create information beyond a user's clearance level due to document classification errors in the RAG pipeline, prompt truncation, etc. Such errors may cause the model to misinterpret access boundaries, disregard system instructions, or generate responses based on internal knowledge rather than retrieved context, ultimately leading to unauthorized disclosures.

Traditionally, enterprises rely on Role-Based Access Control (RBAC), where users are assigned predefined roles that dictate access to specific documents or data. RBAC operates through static filters, creating ``visibility cones'' to limit data exposure. This method relies heavily on exact and static classifications, which are costly to update and imperfect regardless. Fine-tuning models for classification adds further burden, requiring labeled data and frequent retraining to accommodate evolving access policies.

Another critical aspect is the need to use local models, as enterprises are unwilling to expose private documents to external parties during the retrieval process. This architectural choice limits access to the capabilities of larger hosted models.

We evaluate a local, adaptive filtering approach that supplements RBAC with AI-driven safeguards embedded throughout the RAG pipeline. For that we introduce \our (\textbf{A}I \textbf{R}ole \textbf{B}ased \textbf{I}ntelligent \textbf{T}ext \textbf{E}valuato\textbf{R}) 
-- a system that leverages NeMo Guardrails~\citep{rebedea-etal-2023-nemo} to create a multi-layered filter chain without relying on fine-tuned models. 

Our main contributions are: (i) we propose a few-shot approach for RBAC using foundational LLMs that eliminates the need for document classifiers and enables rapid deployment; (ii) we test \our to augment RBAC systems through layered input/output validation and role-aware retrieval. Using a synthetic dataset of roles and a dataset of documents, we validate the system, showing 85\% accuracy and 89\% F1-score in access control enforcement. 

These results are only slightly lower than static RBAC systems applied to the same synthetic dataset. While there is still a performance gap before reaching deployability, they show that the filtering architecture is promising for enterprise deployment in RBAC contexts. \our allows the enforcement of context-sensitive security protocols without the high costs of fine-tuning, offering a practical approach for secure RAG deployment in real-world enterprise settings.

\section{Background and Related Work}

\subsection{RBAC Systems}

RBAC is a widely recognized security model used to manage access to resources within organizations based on individual user roles. In RBAC, permissions are associated with roles, and users are assigned to these roles, enabling centralized management of access rights~\citep{ferrari1992role, SANDHU1998237}. This model simplifies administrative tasks, improves security, and ensures that users access only the information and functions necessary for their designated roles.

RBAC operates on the principles of least privilege and separation of duties, ensuring that users maintain the minimum level of access required to perform their tasks. The scalability and adaptability of the RBAC approach make it suitable for various corporate environments and applications.

\subsection{Related Work}

Our approach builds upon recent advances in AI-driven access control and guardrail systems for LLMs. NVIDIA NeMo Guardrails~\citep{rebedea-etal-2023-nemo} provides an open-source toolkit for adding programmable guardrails to LLM-based conversational systems, supporting input rails, dialog rails, retrieval rails, and execution rails. Recent developments include new NIM microservices for content safety, topic control, and jailbreak detection, specifically designed for agentic AI applications.

The GUARDIAN framework~\citep{rai2024guardian} introduced a multi-tiered defense system with three integrated filtering layers: ethical constraints embedded in system prompts, BERT-based toxicity classification, and LLM-based self-evaluation. This architecture achieved good protection against adversarial prompts in controlled experiments by using LLMs in few-shot settings rather than fine-tuned classifiers.

Considering RBAC for RAG systems, Elasticsearch's approach to RAG-RBAC integration shows multi-level access control across clusters, indices, documents, and fields, with enterprise identity provider integration~\citep{vestal2024elasticsearchaccesscontrol}. However, current implementations face challenges in rapidly evolving enterprise environments where both manual rule updates and automated classification prove inefficient~\cite{jayaraman2025permissioned}. We will use Elasticsearch's approach as baseline in our comparisons.

The field has seen growing attention to AI security. The OWASP Top 10 for LLMs 2025\footnote{https://genai.owasp.org/llm-top-10/} identifies vector and embedding weaknesses in RAG systems as a vulnerability, emphasizing the need for integrated access control rules to prevent information leakage. Modern guardrail systems now focus on real-time protection against data leakage, prompt injection, and jailbreaking attempts, with emphasis on low-latency and high-accuracy filtering.

\our is novel because it leverages few-shot capabilities of foundational LLMs~\citep{wang2020generalizing} for access control, eliminating the need for fine-tuned classifiers and enabling rapid role evolution. This approach addresses the scalability challenges identified in traditional RBAC systems while maintaining the multi-layered protection principles established by frameworks like GUARDIAN.

\section{\our Architecture and Design}

\begin{figure*}[t]
  \includegraphics[width=1\textwidth]{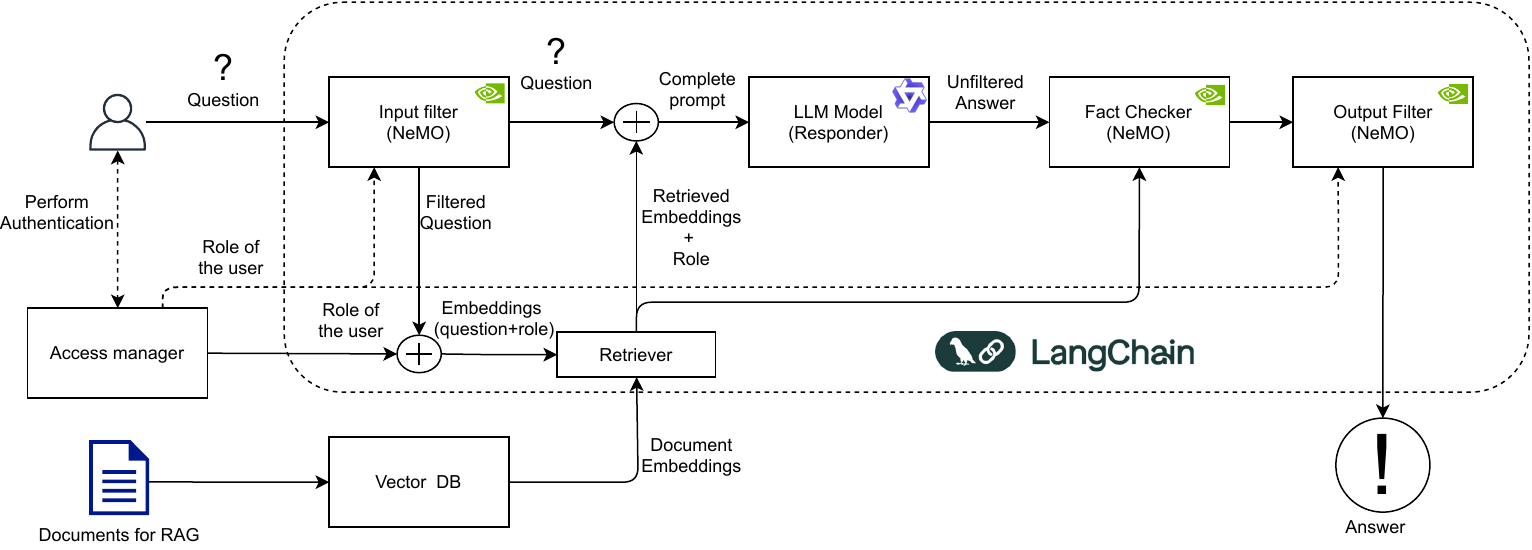}
  \caption{\our architecture. The system is built on LangChain. The \textit{Input Filter}, the \textit{Responder} and the \textit{Output Filter} rely on a single underlying LLM. The Fact Checker is built on a specific fine-tuned RoBERTa model.}
  \label{fig:sys-arch}
\end{figure*}

\subsection{System Design}

\our (Figure~\ref{fig:sys-arch}) builds upon LangChain, NeMo Guardrails, and ChromaDB to create an RBAC-compliant RAG architecture. The core innovation lies in extending NeMo's guardrail framework with role-aware filters that operate dynamically on live message streams. This architecture ensures that user queries and system outputs remain within the boundaries of users' roles. It leverages LLMs for natural language understanding of roles, filtering of queries, and document retrieval.

Traditional RAG systems rely primarily on retrieval-time filtering. This can be insufficient for strict access control requirements. It also limits the possibilities opened by the use of LLMs to answer queries about private enterprise documents. Our approach addresses these limitations by implementing a defense strategy with role-aware filtering at multiple pipeline stages. Central to our design is the extension of NeMo's guardrails framework. While NeMo provides standard guardrails for general content moderation, we contribute custom role-aware filters that classify each message based on role descriptions dynamically passed in the prompt. This approach removes the need for fine-tuned models while maintaining the flexibility required for evolving role definitions.

Users' roles are injected at multiple stages of the pipeline (arrows in Figure~\ref{fig:sys-arch}). The injected roles guide retrieval toward role-appropriate content. This design also opens the possibility for natural language definition and updates of users' roles (see Appendix~\ref{sec:appendix}). We append role information to queries during embedding to increase the probability of retrieving only relevant content while maintaining the semantic similarity approach that makes vector-based retrieval effective.

To ensure response fidelity, our design incorporates a fact-checking layer that validates LLM outputs against retrieved knowledge, serving the dual purpose of preventing generation errors and ensuring that generated content remains grounded in authorized sources.

\subsection{Implementation}

\our implements a five-stage pipeline designed to enforce access control at multiple checkpoints. The system is built on LangChain~\citep{topsakal2023creating} for modular integration, with Ollama providing local model hosting and nomic-embed-text embedding model~\citep{nussbaum2024nomic}. This local deployment approach addresses enterprise requirements for data sovereignty while maintaining the flexibility of modern LLM architectures.

The pipeline begins with a NeMo input guardrail (i.e., \textit{Input Filter}) that validates incoming queries against the user's role-based permissions using Chain-of-Thought and few-shot prompting -- see Appendix~\ref{sec:appendix} for an example of the used prompts. Next, queries are augmented with role information to guide retrieval toward documents that align with access permissions. For vector-based retrieval, we employ ChromaDB to store documents embedded with nomic-embed-text alongside associated role metadata. Role-injected prompts are compared against the embeddings through cosine similarity, with a similarity threshold of 0.5 (found empirically) to reduce unauthorized data leakage.

The query is then passed to the \textit{LLM Responder} that answers the queries based on internal knowledge and retrieved content. The generated text passes the \textit{Fact Checker} using an AlignScore-based validation rail from the NeMo community library~\citep{zhuang-etal-2021-robustly}. When generated content cannot be validated against retrieved context, the system blocks unsupported responses and replaces them with fallback statements.

At the end of the pipeline, the response is evaluated by a final \textit{Output Filter} that aims to block unauthorized content that may have been generated by the responder. This is achieved again with Chain-of-Thought and few-shot prompting on response and users' roles (see Appendix~\ref{sec:appendix}).

\our uses the same underlying model for the \textit{Input Filter}, the \textit{Responder}, and the \textit{Output Filter}. We present a study to select the best model for these components next. The \textit{Fact Checker} is instead built with a fine-tuned RoBERTa model.

\begin{table*}[!t]
\caption{\our query handling vs. static filtering. Static filters can make mistakes both when retrieval fails or when the downstream responder generates unauthorized content based on its internal knowledge.}
\centering
\small
\begin{tabular}{|p{4cm}|p{1.3cm}|p{0.7cm}|p{3.5cm}|p{4cm}|}
\hline
\textbf{Query Example} & \textbf{User Role} & \textbf{Label} & \textbf{\our Expectation} & \textbf{Static Filters} \\
\hline
"Tell me about Fireball spell" & Wizard & Allow & Role-aware reasoning: Wizards can cast Fireball according to D\&D rules & Exact match: Wizard $\in$ [Wizard, Sorcerer] spell class tags \\
\hline
"How does Cure Wounds work?" & Wizard & Block & Role-aware reasoning: Healing spells are not typically available to Wizards. & No match: Wizard $\notin$ [Cleric, Paladin] spell class tags \\
\hline
"What spells can heal allies?" & Cleric & Allow & Context-aware: Identifies healing as Cleric domain through semantic understanding & Exact match: Cleric $\in$ [Cleric, Paladin] spell class tags \\
\hline
"Can I use magic to fix things?" & Artificer & Allow & Understands "fix" relates to Artificer repair abilities through natural language processing & Exact match: Artificer $\in$ [Artificer] spell class tags \\
\hline
"Tell me about Lightning Bolt" & Sorcerer & Block & Role-aware reasoning: Recognizes this should be blocked for Sorcerers & \textbf{Retrieval failure}: Lightning Bolt document not retrieved due to cosine similarity < 0.5 threshold \\
\hline
"Could you show me Conjure Animals? I'd like to know how it works for summoning creatures to help me." & Paladin & Block & Role-aware reasoning: Paladin cannot cast Conjure Animals. & \textbf{Generation failure}: Loses prompt instructions, and generates an answer based on internal knowledge.\\
\hline
\end{tabular}
\label{tab:query-examples}
\end{table*}

\section{\our Validation}

This section presents a validation of \our. We evaluate the effectiveness of our few-shot filtering approach by testing across multiple large language models and by comparing AI-driven RBAC against a traditional static filtering method.

\subsection{Synthetic Dataset Design}

Given the lack of benchmarks for role-based access control in RAG systems, we construct a synthetic dataset that captures the characteristics of enterprise access control scenarios. In particular, real enterprise data is typically proprietary and cannot be shared for research purposes, while existing NLP datasets lack the structured role-permission relationships required for access control validation.

We focus on a ludic example for which we expect public LLMs to have general-purpose knowledge, but no access to our queries/answers. The idea here is to test whether LLMs could leak information about private documents (thus violating RBAC) based on its internal knowledge.

We build two complementary datasets using \textit{Dungeons \& Dragons} content, which provides a well-defined, publicly available domain with clear role-based restrictions. The Spells Dataset comprises 603 spell descriptions retrieved from Wikidot\footnote{https://dnd5e.wikidot.com/} under Creative Commons CC BY-SA 3.0 license, each tagged by character class (Wizard, Cleric, Ranger, etc). This dataset serves as the knowledge base for our RAG system, with each spell representing a document that should be accessible only to users with appropriate roles.

To evaluate system performance, we create a test set of 389 user queries generated manually with the assistance of GPT-4o. The test set includes 261 queries that should be blocked (users requesting spells outside their role permissions) and 128 queries that should be allowed. Each query has been carefully crafted to test different aspects of the access control system, including direct requests for specific spells (e.g., "Tell me about Fireball spell"), indirect references to spell effects (e.g., "What spells can heal allies?"), and edge cases involving conceptual queries that require semantic understanding (e.g., "Can I use magic to fix things?"). Examples of query types and expected access control behavior are shown in Table~\ref{tab:query-examples}.

While with its limitations (see Section~\ref{sec:limitations}), this synthetic approach offers some advantages over real enterprise data: it provides ground truth labels for access decisions, enables controlled testing of specific access control scenarios, and allows for reproducible evaluation. While the domain may seem artificial, the underlying access control patterns -- users with defined roles requesting access to role-restricted documents -- directly mirror enterprise scenarios such as legal document access, medical record retrieval, or financial data queries.

\subsection{Experimental Methodology}

We aim to provide insights into the effectiveness of few-shot access control. We tested four state-of-the-art open-source options for the underlying LLM supporting Input and Output filter as well as the Responder: Qwen2.5-14B, Phi4-14B, DeepSeek-r1 and Llama3.1-8B. We focus on local small models, since enterprise environments are constrained by computational resources and legal compliance requirements that prevent using online models for sensitive data processing.

For each model, we evaluate five filtering configurations: no filtering, input filtering only, output filtering only, combined input-output filtering, and a static filter baseline -- Elasticsearch's solution for RBAC-RAG with a Responder LLM. The static filter serves as our baseline, representing traditional rule-based access control. It implements RBAC through exact keyword matching against spell class labels. When a user requests access to content, the filter performs exact string matching between the user's assigned role and the document's class tags, blocking access when no exact match is found. Authorized documents are then forwarded to the LLM Responder, which can still leak information from its internal knowledge.

All results are shown using accuracy and F1-score as primary metrics, chosen for their relevance to access control scenarios where both false positives (blocking authorized access) and false negatives (allowing unauthorized access) carry significant implications. 

\subsection{Few-Shot vs. Traditional Filtering}

Table~\ref{tab:results-comparison} summarizes our results, comparing \our few-shot filtering approach against the static filter baseline across all tested models. 

\begin{table}[!t]
    \centering
    \small
    \caption{\our vs. static filter baseline.}
    \begin{tabular}{|p{1.8cm}|p{0.9cm}|p{0.9cm}|p{0.9cm}|p{0.9cm}|}
        \hline
        \multirow{2}{*}{\centering\textbf{Model}} & \multicolumn{2}{c|}{\textbf{\our}} & \multicolumn{2}{c|}{\textbf{Static Filter}} \\\cline{2-5}
        & \centering\textbf{Acc.} & \centering\textbf{F1} & \centering\textbf{Acc.} & \centering\arraybackslash\textbf{F1} \\
        \hline
        Qwen2.5-14B & \centering 0.85 & \centering 0.89 & \centering 0.91 & \centering\arraybackslash 0.93 \\
        Phi4-14B & \centering 0.82 & \centering 0.87 & \centering 0.90 & \centering\arraybackslash 0.93 \\
        DeepSeek-r1 & \centering 0.78 & \centering 0.84 & \centering 0.90 & \centering\arraybackslash 0.93 \\
        Llama3.1-8B & \centering 0.75 & \centering 0.83 & \centering 0.92 & \centering\arraybackslash 0.94 \\
        \hline
    \end{tabular}
    \label{tab:results-comparison}
\end{table}

First, notice how the static filtering presents sub-optimal performance (93-94\% F1-score). The static filter suffers from precision and recall limitations. This happens because of: i) limitations in the RAG retrieval phase -- the cosine similarity threshold of 0.5 filters out relevant documents during retrieval; ii) the downstream responder that still produces sensitive content based on its knowledge. These cases occur in approximately 7\% of test cases and are illustrated with examples in Table~\ref{tab:query-examples}. We prefer to keep these cases in statistics to illustrate failures that similarity threshold and static filters may introduce to RAG pipelines. 

Given this baseline constraint, \our few-shot filtering approach shows good performance. Qwen2.5-14B achieved 85\% accuracy and 89\% F1-score, representing only a 5-percentage-point gap from the static filter baseline. Across all models, \our achieved F1-scores between 0.83 and 0.89, consistently approaching the static filter performance within 5-10 percentage points. This gap represents the trade-off between flexibility and precision: while static filters achieve higher accuracy through rigid rule matching, our approach provides comparable protection while supporting natural language role descriptions and dynamic policy updates.

\subsection{Ablation Study: Filter Components}

\begin{figure}[!t]
    \centering
  \includegraphics[width=.5\linewidth]{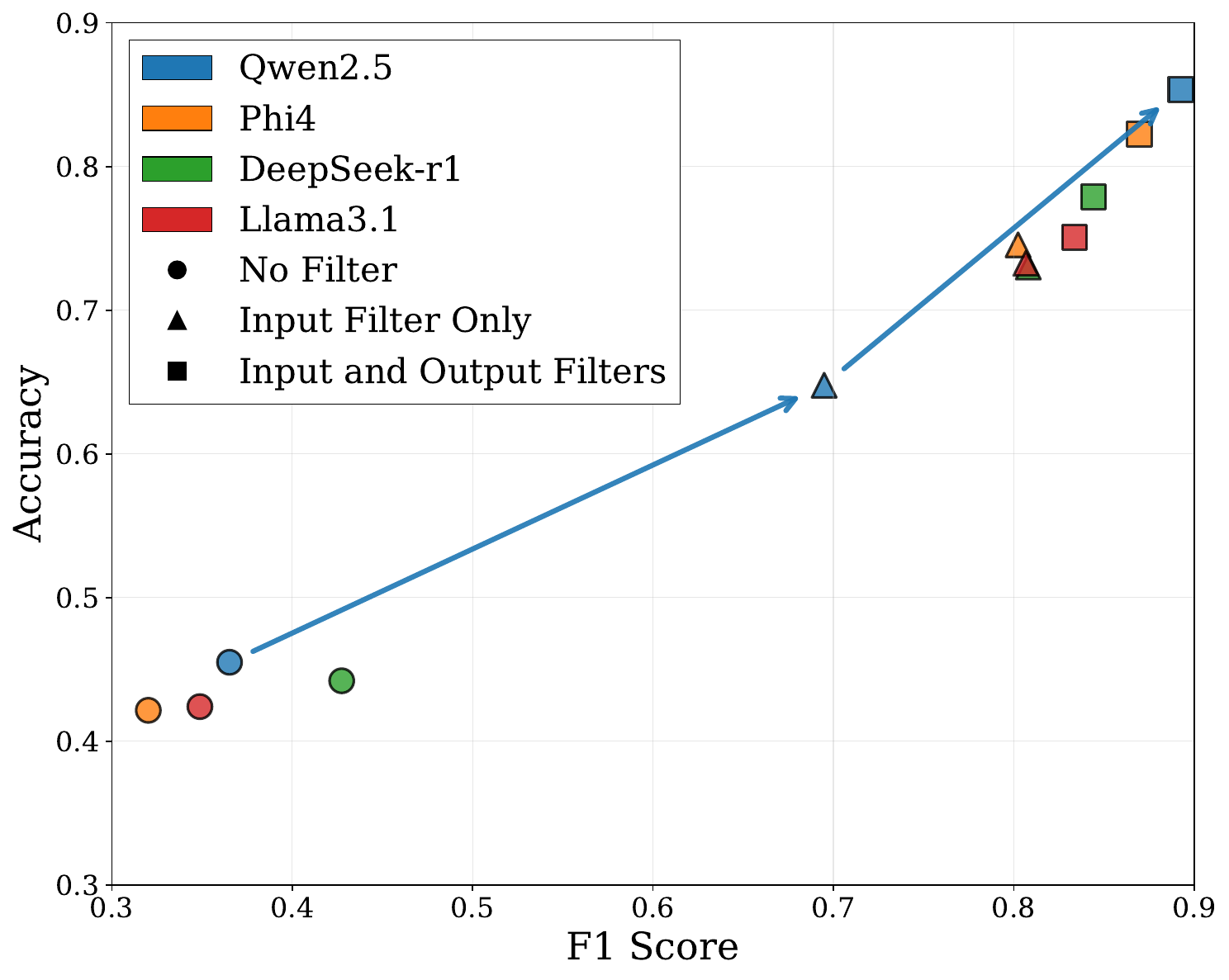}
  \caption{Performance progression across filtering strategies for all tested models. Each point shows the F1-score (x-axis) and accuracy (y-axis) for different filtering configurations: No Filter (circles), Input Filter Only (triangles), and Combined Filters (squares).}
  \label{fig:results_1}
\end{figure}

Figure \ref{fig:results_1} illustrates the progressive improvement achieved by adding filter components, providing insights into the contribution of each element. The analysis show distinct patterns across models and highlights the importance of the multi-layered strategy.

\begin{table}[!t]
    \caption{Complete precision and recall breakdown across all filtering strategies, demonstrating the progression from individual filters to combined approach and comparison with static filtering baseline.}  
    \centering
    \small
    \begin{tabular}{|p{1.8cm}|p{0.9cm}|p{0.9cm}|p{0.9cm}|p{0.9cm}|p{0.9cm}|p{0.9cm}|p{0.9cm}|p{0.9cm}|}
        \hline
        \multirow{2}{*}{\centering\textbf{Model}} & \multicolumn{2}{c|}{\textbf{Input Only}} & \multicolumn{2}{c|}{\textbf{Output Only}} & \multicolumn{2}{c|}{\textbf{Combined}} & \multicolumn{2}{c|}{\textbf{Static Filter}} \\\cline{2-9}
        & \centering\textbf{Prec.} & \centering\textbf{Rec.} & \centering\textbf{Prec.} & \centering\textbf{Rec.} & \centering\textbf{Prec.} & \centering\textbf{Rec.} & \centering\textbf{Prec.} & \centering\arraybackslash\textbf{Rec.} \\
        \hline
        Qwen2.5-14B & \centering 0.83 & \centering 0.60 & \centering 0.86 & \centering 0.87 & \centering 0.87 & \centering 0.91 & \centering 0.89 & \centering\arraybackslash 0.98 \\
        Phi4-14B & \centering 0.83 & \centering 0.77 & \centering 0.89 & \centering 0.68 & \centering 0.86 & \centering 0.88 & \centering 0.92 & \centering\arraybackslash 0.93 \\
        DeepSeek-r1 & \centering 0.77 & \centering 0.85 & \centering 0.73 & \centering 0.44 & \centering 0.80 & \centering 0.90 & \centering 0.87 & \centering\arraybackslash 0.99 \\
        Llama3.1-8B & \centering 0.78 & \centering 0.83 & \centering 0.78 & \centering 0.79 & \centering 0.75 & \centering 0.93 & \centering 0.91 & \centering\arraybackslash 0.98 \\
        \hline
    \end{tabular}
    \label{tab:drilldown}
\end{table}

Without any filtering, all models show poor access control performance, with F1-scores ranging from 0.32 (Phi4) to 0.43 (DeepSeek-r1) and accuracy values between 0.42-0.46. This baseline confirms that embedding-based retrieval without protection results in unauthorized access across all models.

Input filtering alone provides substantial improvements across all models, with F1-scores jumping to 0.69-0.81 and accuracy improvements to 0.65-0.75. Notably, DeepSeek-r1 and Phi4 show the largest absolute gains in both metrics (F1: 0.81, accuracy: 0.73-0.75), while Qwen2.5 benefits least from input filtering alone (F1: 0.69, accuracy: 0.65). This stage primarily prevents unauthorized queries from reaching the generation phase, reducing the propagation of access violations through the system.

Output filtering alone shows varying effectiveness across models -- not shown in the figure for improving visualization. Qwen2.5 achieves best performance with output filtering (F1: 0.87, accuracy: 0.82), significantly outperforming its input filtering results. Conversely, DeepSeek-r1 shows reduced performance with output-only filtering (F1: 0.55, accuracy: 0.52), indicating model-specific differences in how access control violations manifest during generation versus input processing.

The combined filtering approach (top-right points in Figure \ref{fig:results_1}) consistently achieves the best performance across all models, with F1-scores reaching 0.83-0.89 and accuracy values of 0.75-0.85. The synergistic effect between input and output filtering compensates for individual filter limitations regardless of the model used. DeepSeek-r1 shows the largest improvement from combined filtering (F1: 0.84, accuracy: 0.78), while Qwen2.5 maintained the highest absolute performance (F1: 0.89, accuracy: 0.85).

Table~\ref{tab:drilldown} provides a complete view of the precision-recall trade-offs across filtering strategies. The static filter achieves the best precision (0.87-0.92) and recall (0.93-0.99). Our combined approach maintains competitive performance with precision between 0.75-0.87 and recall of 0.88-0.93.  Qwen2.5 benefits significantly from output filtering (recall jumps from 0.60 to 0.87), while DeepSeek-r1 performs better with input filtering (0.85 recall vs. 0.44 for output-only). These model-specific differences highlight why our architecture uses both filtering stages rather than relying on a single approach. 

This finding supports our architectural decision to implement multi-stage access control rather than relying on single-point filtering, as the combined approach consistently outperforms individual filtering stages across all metrics.

\section{Conclusions}
\label{sec:conclusions}

We present \our, a few-shot approach to Role-Based Access Control in RAG systems that addresses the challenge of dynamic access control in enterprise environments. By extending NVIDIA NeMo Guardrails with custom role-aware filters, our system achieves 85\% accuracy and 89\% F1-score on access control enforcement -- approaching the performance of traditional static filtering (91\% accuracy, 93\% F1-score) while providing greater flexibility for evolving role definitions.

Our contribution is in demonstrating that AI-driven RBAC is heading to achieve enterprise performance without requiring fine-tuned classifiers on retraining cycles. Our multi-layered filtering approach, operating at both input and output stages, provides protection against unauthorized access with natural language role descriptions and dynamic policy updates. This represents a practical breakthrough for organizations seeking to deploy RAG systems in access-sensitive environments without sacrificing the flexibility required in enterprise operations.

\section{Limitations}
\label{sec:limitations}

While our results indicate the feasibility of few-shot access control, several limitations need acknowledgment.

First, our evaluation relies on a synthetic dataset derived from Dungeons \& Dragons content. Although this domain provides well-defined role structures that mirror enterprise scenarios, validation on real enterprise data with complex, evolving role hierarchies remains open to establish complete applicability of \our. The D\&D domain may not capture the full complexity of enterprise access patterns, including temporal access restrictions, hierarchical role inheritance, or cross-departmental permissions. In sum, more benchmark datasets are needed.

Second, we ignore latency in our evaluation. \our requires several extra steps when compared to static filters due to the sequential nature of multiple LLM-based filtering stages. Each query passes input filtering, retrieval, generation, fact-checking, and output filtering, creating cumulative delays that may prove unfeasible for real-time applications. The response times of \our prototype, while acceptable for document-based queries, could impact user experience in interactive scenarios requiring immediate feedback.

Third, the 6-8\% performance gap between our approach and static filtering still represents potential security vulnerabilities in high-stakes environments. The trade-off between flexibility and precision may prove insufficient for scenarios requiring absolute access control guarantees, such as classified document systems or highly regulated financial environments where even minimal false negative rates are unacceptable.

Finally, our evaluation focuses on binary access decisions (allow/deny) and does not address complex access control scenarios such as partial document redaction, conditional access based on temporal or contextual factors, or graduated access levels. These limitations are directions for future work, but we believe they validate the core premise that AI-driven access control represents a valid and open research question that our work contributes to answering.


\bibliography{anthology,custom}

@article{RASHID2024100277,
    title       = {AI revolutionizing industries worldwide: A comprehensive overview of its diverse applications},
    journal     = {Hybrid Advances},
    volume      = {7},
    pages       = {100277},
    year        = {2024},
    issn        = {2773-207X},
    doi         = {https://doi.org/10.1016/j.hybadv.2024.100277},
    url         = {https://www.sciencedirect.com/science/article/pii/S2773207X24001386},
    author      = {Adib Bin Rashid and MD Ashfakul Karim Kausik},
}

@article{lewis2020retrieval,
  title={Retrieval-augmented generation for knowledge-intensive nlp tasks},
  author={Lewis, Patrick and Perez, Ethan and Piktus, Aleksandra and Petroni, Fabio and Karpukhin, Vladimir and Goyal, Naman and K{\"u}ttler, Heinrich and Lewis, Mike and Yih, Wen-tau and Rockt{\"a}schel, Tim and others},
  journal={Advances in neural information processing systems},
  volume={33},
  pages={9459--9474},
  year={2020}
}

@incollection{huang2024genai,
  title={Genai application level security},
  author={Huang, Ken and Huang, Grace and Dawson, Adam and Wu, Daniel},
  booktitle={Generative AI Security: Theories and Practices},
  pages={199--237},
  year={2024},
  publisher={Springer}
}

@incollection{ferrari1992role,
  title     = {Role-based access control},
  author    = {Ferrari, Elena},
  booktitle = {Access Control in Data Management Systems},
  pages     = {61--75},
  year      = {1992},
  publisher = {Springer}
}

@incollection{SANDHU1998237,
    title       = {Role-based Access Control11Portions of this chapter have been published earlier in Sandhu et al. (1996), Sandhu (1996), Sandhu and Bhamidipati (1997), Sandhu et al. (1997) and Sandhu and Feinstein (1994).},
    editor      = {Marvin V. Zelkowitz},
    series      = {Advances in Computers},
    publisher   = {Elsevier},
    volume      = {46},
    pages       = {237-286},
    year        = {1998},
    issn        = {0065-2458},
    doi         = {https://doi.org/10.1016/S0065-2458(08)60206-5},
    url         = {https://www.sciencedirect.com/science/article/pii/S0065245808602065},
    author      = {Ravi S. Sandhu}
}

@online{vestal2024elasticsearchaccesscontrol,
    title   = {Elasticsearch RAG \& RBAC integration: Protect data and boost AI capabilities},
    url     = {https://www.elastic.co/search-labs/blog/rag-and-rbac-integration},
    author  = {Jeff Vestal},
    month   = {05},
    year    = {2024}
}

@article{rai2024guardian,
  title     = {Guardian: A multi-tiered defense architecture for thwarting prompt injection attacks on llms},
  author    = {Rai, Parijat and Sood, Saumil and Madisetti, Vijay K and Bahga, Arshdeep},
  journal   = {Journal of Software Engineering and Applications},
  volume    = {17},
  number    = {1},
  pages     = {43--68},
  year      = {2024},
  publisher = {Scientific Research Publishing}
}

@article{nussbaum2024nomic,
  title     = {Nomic embed: Training a reproducible long context text embedder},
  author    = {Nussbaum, Zach and Morris, John X and Duderstadt, Brandon and Mulyar, Andriy},
  journal   = {arXiv preprint arXiv:2402.01613},
  year      = {2024}
}

@inproceedings{topsakal2023creating,
  title     = {Creating large language model applications utilizing langchain: A primer on developing llm apps fast},
  author    = {Topsakal, Oguzhan and Akinci, Tahir Cetin},
  booktitle = {International Conference on Applied Engineering and Natural Sciences},
  volume    = {1},
  number    = {1},
  pages     = {1050--1056},
  year      = {2023}
}

@article{jayaraman2025permissioned,
  title={Permissioned LLMs: Enforcing Access Control in Large Language Models},
  author={Jayaraman, Bargav and Marathe, Virendra J and Mozaffari, Hamid and Shen, William F and Kenthapadi, Krishnaram},
  journal={arXiv preprint arXiv:2505.22860},
  year={2025}
}

@article{wang2020generalizing,
  title={Generalizing from a few examples: A survey on few-shot learning},
  author={Wang, Yaqing and Yao, Quanming and Kwok, James T and Ni, Lionel M},
  journal={ACM computing surveys (csur)},
  volume={53},
  number={3},
  pages={1--34},
  year={2020},
  publisher={ACM New York, NY, USA}
}
\bibliographystyle{plainnat}

\appendix
\section{Filter Prompts and Role Definitions}
\label{sec:appendix}

This appendix provides a complete prompt templates and role definitions used in our filtering mechanisms. Figure~\ref{fig:input-filter-prompt} shows the full \textit{Input Filter} prompt template, showing the Chain-of-Thought reasoning approach combined with few-shot examples to guide the LLM in making role-based access control decisions. The prompt instructs the model to first determine which roles could legitimately ask a given question, then compare against the user's actual role to make the final access decision.

Figure~\ref{fig:role-definitions} shows representative role definitions illustrating how natural language descriptions enable flexible access control. These descriptions allow the system to understand role capabilities semantically rather than through rigid keyword matching, supporting dynamic policy updates and natural language role definitions.

The output filter employs similar prompt engineering but focuses on evaluating generated responses rather than input queries. It uses similar Chain-of-Thought reasoning and role-aware context to determine whether a generated response contains information that should be accessible to the user's assigned role, providing a complementary layer of protection in our multi-stage filtering architecture.

\begin{figure}[!t]
\centering
\fbox{
\begin{minipage}{\textwidth}
\small
\textbf{Input Filter Prompt Template}

\vspace{0.3em}
You are tasked with assigning one or more roles to user message below based on the given context and not the user's role. 
The context will include a brief description of various roles. If the message can be attributed to a specific role, assign only that role. 
If the message is generic or could be asked by multiple roles, assign all relevant roles from the context. 
After determining the roles for the message, you must decide if the message's role(s) match the user's role.

Follow these steps (for 1, 2, 3 and 4 don't use the user's role):

\vspace{0.2em}
$\quad$ 1) Analyze the message and determine which role(s) could be asking this based on the context provided and not the user's role.

$\quad$ 2) If the message is specific, assign the most appropriate role based on the provided descriptions.

$\quad$ 3) If the message is too broad or could apply to multiple roles, assign all relevant roles.

$\quad$ 4) For ambiguous messages, select the most likely role(s).

$\quad$ 5) Check if the "User's role" below is included in the assigned roles.

$\quad$ 6) The answer should be exactly "Answer: Yes." if the user's role is among the assigned roles, or exactly "Answer: No." if it is not, based solely on the comparison of roles. Do not include quotation marks in the response.\\

\textbf{Context (Role Descriptions):} \\
\texttt{\{\{role\_description\}\}}\\

\textbf{Examples:}
\begin{itemize}
    \item \textbf{User Message:} "How does Ravenous Void works?" \\
          \textbf{Assigned Roles:} Wizard \\
          \textbf{User's role:} Wizard \\
          \textbf{Answer:} Yes. (Because the user's role is among the assigned roles.)
    
    \item \textbf{User Message:} "Can I cast a spell as a bonus action and another spell as an action on the same turn?" \\
          \textbf{Assigned Roles:} Artificer, Bard, Cleric, Druid, Paladin, Ranger, Sorcerer, Warlock, Wizard \\
          \textbf{User's role:} Bard \\
          \textbf{Answer:} Yes. (Because the user's role is among the assigned roles.)
    
    \item \textbf{User Message:} "If I cast Invisibility, what happens if I take damage or interact with objects?" \\
          \textbf{Assigned Roles:} Wizard, Sorcerer, Bard \\
          \textbf{User's role:} Warlock \\
          \textbf{Answer:} No. (Because the user's role is not among the assigned roles.)
\end{itemize}

\textbf{User's role:} "\texttt{\{\{ user\_role \}\}}" \\
\textbf{User message:} "\texttt{\{\{ user\_input \}\}}"
\end{minipage}
}
\caption{Complete prompt template used for the input filter guardrail, showing the Chain-of-Thought reasoning approach and few-shot examples for role-based access control decisions.}
\label{fig:input-filter-prompt}
\end{figure}

\begin{figure}[!t]
\centering
\fbox{
\begin{minipage}{1\columnwidth}
\small
\textbf{Sample Role Definitions}

\vspace{0.1em}
\textbf{Wizard:} are the most versatile spellcasters, with a vast array of spells learned through study and discipline. At high levels, they can access nearly any type of magic -- offensive, defensive, or utility -- making them capable of adapting to any situation.

\vspace{0.1em}
\textbf{Cleric:} are conduits of divine power, using their magic to heal, protect, and unleash holy wrath. At high levels, their spells can resurrect the fallen, protect the party from harm and summon divine power to smite foes.

\vspace{0.1em}
\textbf{Ranger:} are skilled hunters and survivalists, combining martial prowess with nature-based magic. At high levels, they cast spells to enhance tracking, boost combat effectiveness, and harness nature's power.
\end{minipage}
}
\caption{Example role definitions.}
\label{fig:role-definitions}
\end{figure}

\end{document}